\documentclass[epj]{svjour}
\usepackage{graphics}
\usepackage{amssymb}
\usepackage{amsmath}
\newcommand{\half}{\mbox{${\textstyle \frac{1}{2}}$}}           % 1/2
\newcommand{\fourth}{\mbox{${\textstyle \frac{1}{4}}$}}         % 1/4
\newcommand{\fmn}[2]{\mbox{${\textstyle \frac{#1}{#2}}$}}

\newcommand{\bmath}[1]{\mbox{\boldmath $#1$}}
\newcommand{\fsi}{\emph{fsi}}
\newcommand{\RK}{\mbox{$\mathcal{R}(K^+)$}}

\begin{document}
\title{Spin and Isospin Effects in the $NN\rightarrow NK\Lambda$
Reaction Near Threshold}

\author{G\"{o}ran F\"{a}ldt\inst{1} \and Colin Wilkin\inst{2}}

\institute{
Department of Radiation Sciences, Box 535, S-751 21 Uppsala, Sweden,
\email{goran.faldt@tsl.uu.se}
\and
Department of Physics \& Astronomy, UCL, London WC1E 6BT, UK,
\email{cw@hep.ucl.ac.uk}}

\date{Received: \today / Revised version: date}

\abstract{The spin and isospin structure of the amplitudes and
observables for $K^+\Lambda$ production in nucleon-nucleon
collisions in the near-threshold region is analysed. It is shown
that, with reasonable values for the relative strengths of the
$\pi$ and $\rho$ terms in a meson-exchange model, one expects
production on the neutron to be significantly stronger than that
on the proton. Negative values of the spin-transfer coefficient
$D_{NN}$ are also predicted due to $\pi$-$\rho$ interference.
}
\maketitle
%
%%%%%%%%%%%%%%%%%%%%%%%%%%%%%%%%%%%%%%%%%%%%%%%%%%%%%%%%%%%%%%%%%%%%%%
%
\section{Introduction}
The study of meson production near threshold in nucleon-nucleon
scattering has been a growth area over the last decade and most of
the modelling of experimental data has been in terms of some form
of meson-exchange model~\cite{review,Hanhart}. The energy
dependence of the total cross section is generally dominated by
phase-space, folded with a strong nucleon-nucleon final-state
interaction (\fsi)~\cite{FW0}. The information that the data give
on the basic driving term is therefore very limited and has to be
supplemented by results from angular distributions and Dalitz
plots \emph{etc}. A particularly valuable constraint on
theoretical models is the relative strength of the production in
neutron-proton and proton-proton collisions. For $\eta$
production, it is found that %
$\mathcal{R}(\eta)=\sigma(pn\to\eta pn)/\sigma(pp\to\eta pp) =
6.5\pm 1.0$~\cite{Calen}. Neglecting the differences between the
$np$ and $pp$ initial and final-state interactions, the exchange
of just a single pion or $\rho$ meson would lead to a factor of
5~\cite{GW}, which is close to the experimental value. This is
reduced by the \fsi, but a quantitative agreement with the
observables can be obtained with $\rho$-meson exchange being more
important than that of the $\pi$~\cite{FW1}, though alternative
scenarios are in the literature~\cite{Speth1,Speth2}.

The COSY11~\cite{COSY11,Kowina} and
COSY-TOF~\cite{COSY-TOF,COSY-TOF2} collaborations have made
measurements in proton-proton collisions of both $K^+\Lambda$ and
$K\Sigma$ production near their respective production thresholds.
The excitation functions look broadly similar to those for
reactions such as $pp\to pp\eta$, though the effects of the final
state interaction are somewhat less because the hyperon-nucleon
scattering lengths are much smaller than that in $pp$. Though
proton-neutron data are much more sparse, there are strong
indications from reactions on deuterium that \RK\ is also
significantly over unity~\cite{Valdau}. It is the purpose of this
note to explore whether a large value of \RK\ could be understood
within a meson exchange model.

The $pp\to pK^+\Lambda$ cross section near threshold has been
estimated by several groups~\cite{Laget,FW2,Tsushima,Shyam,Speth},
but there has been no general consensus as to whether the reaction
is driven mainly by the exchange of strange or non-strange mesons.
In part this is due to the tremendous uncertainty in the
$pK\Lambda$ coupling constant, as well as in the off-shell
behaviour of the $K^+p$ scattering amplitude, which is not
resonance dominated. However, the recent results from
COSY-TOF~\cite{COSY-TOF2} clearly indicate that the Dalitz plots
for $pp\to pK^+\Lambda$ are dominated by the excitation of nucleon
isobars, though modified by the $\Lambda p$ \fsi. At their lowest
beam momentum (2.85~GeV/c) only the $N^*(1650)$ was seen but the
$N^*(1710)$ becomes steadily more important as the momentum is
raised. This suggests that strange meson exchange, which cannot
excite such isobars, plays only a minor role in the process.
Nevertheless, it is impossible as yet to estimate reliably the
overall rate for the reaction, principally because of the
uncertainty in the final $\Lambda p$ wave function, especially at
short distances. Different modern potentials that reproduce the
limited scattering data give values of the singlet scattering
length that vary from $-0.71$~fm to $-2.51$~fm~\cite{Nijmegen}. We
therefore limit ourselves to the evaluation of cross section
ratios, which depend far less on the distortion in the final, or
initial state~\cite{Speth2}.

Only three amplitudes are necessary to describe $NN\to K\Lambda N$
near threshold, and their spin and isospin structure are
identified in sect.~2, where cross sections and spin observables
are written in terms of them. Since there are more than five
possible observables, there must be relations between some of
them, and one of these is illustrated here. The contributions to
the spin-isospin amplitudes are studied in sect.~3 within a
meson-exchange model. Though strange and non-strange exchanges are
considered, detailed evaluation is confined to the case where only
the $\pi$ and $\rho$ are important. As discussed in sect.~4, the
energy dependence of the total cross section is determined by the
low energy $\Lambda p$ scattering parameters but the normalisation
depends also upon the $\Lambda p$ interaction at short distances,
which is largely unknown. The variation of \RK\ with the
$\pi/\rho$ strength is shown in sect.~5. With the value scaled
from that used in the $\eta$ case, significantly more $K^+\Lambda$
production is to be expected in $pn$ than in $pp$ collisions.
Furthermore, the spin-transfer parameter might be large and
negative through $\pi$-$\rho$ interference, though neither $\pi$
nor $\rho$ alone lead to a negative value. Our conclusions are
reported in sect.~6.
%
%%%%%%%%%%%%%%%%%%%%%%%%%%%%%%%%%%%%%%%%%%%%%%%%%%%%%%%%%%%%%%%%%%%%%%
%
\section{Amplitudes and observables}
\setcounter{equation}{0}

The most general structure of the isotriplet and singlet
$NN\rightarrow NK\Lambda$ amplitudes near threshold is
\[
\mathcal{M}_1= \left[W_{1,s}\, {\eta}_f^{\,\dagger}\,
\bmath{\hat{p}} \cdot \bmath{\epsilon}_i +
iW_{1,t}\,\bmath{\hat{p}}\cdot ( \bmath{\epsilon}_i \times
             \bmath{\epsilon}_f^{\,\dagger}) \right]\,
\bmath{\chi}_f^{\,\,\dagger}\cdot\bmath{\chi}_i\:,
\]\vspace{-10mm}

\begin{equation}
\mathcal{M}_0= W_{0,t}\,
   \bmath{\hat{p}} \cdot \bmath{\epsilon}_f^{\,\,\dagger}\,{\eta}_i
\ \phi_f^{\,\dagger}\,\phi_i \:,
\label{L_1}
\end{equation}
where \bmath{p} is the incident cm beam momentum. The initial
(final) baryons couple to spin-1 or spin-0, represented by
$\bmath{\epsilon}_i$ ($\bmath{\epsilon}_f$) and ${\eta}_i$
(${\eta}_f$) respectively. Similarly, the $\bmath{\chi}_i$
($\bmath{\chi}_f$) and $\phi_i$ ($\phi_f$) describe the isospin-1
and isospin-0 combinations of the initial $NN$ (final $KN$)
states. It is important to note that, due to the Pauli principle,
$W_{1,t}=0$ for the analogous $pp\to pp\eta$ reaction and that
this vanishing leads to quite different spin and isospin effects
for $K$ and $\eta$ production.

After a little spin algebra, it is seen that the unpolarised
intensities are given by
\[I(pp\!\to\! pK^+\Lambda)=\fourth\sum_{\text{spins}} \mid \langle
f\,|\mathcal{M}_1|\,i\rangle\mid^2\]\vspace{-3mm}%
\begin{equation}\label{Ipp}
=\fourth \left(\mid W_{1,s}\mid^2+ 2\mid W_{1,t}\mid^2\right),
\end{equation}
\[I(pn\!\to\! nK^+\Lambda)=I(pn\!\to\! pK^0\Lambda)\]\vspace{-3mm}%
\begin{equation}
=\fmn{1}{16}\left(\mid W_{1,s}\mid^2+ 2\mid W_{1,t}\mid^2+ \mid
W_{0,t}\mid^2\right), \label{Ipn}
\end{equation}
where there is no interference between the two isospin amplitudes
due to the spin averaging.

One may expect that, close to threshold, the amplitudes
$W_{i,s/t}$ should vary little, except for the different $\Lambda
N$ final-state interactions in the spin-singlet ($s$) and -triplet
($t$) systems. If we neglect these \fsi, the corresponding total
cross section becomes
\begin{eqnarray}
\nonumber \sigma(pp\!\to\! pK^+\Lambda)&=& \frac{1}{64\pi^2ps}
\frac{\left(m_pm_{\Lambda}m_{K}\right)^{1/2}}
{\left(m_p+m_{\Lambda}+m_{K}\right)^{1/2}}\\
&&\times Q^2\, I(pp\!\to\! pK^+\Lambda)\:, \label{sigma}
\end{eqnarray}
and similarly for the $pn$ reaction. Here the $m_i$ are the masses
in the final state, $p$ is the incident proton cm momentum,
$\sqrt{s}$ the total cm energy, and $Q=\sqrt{s}-\sum m_i$, the
excess energy.

In the near-threshold region, both the proton analysing power and
the $\Lambda$ polarisation should vanish and it is only tensor
combinations that are predicted to be non-zero. Of these, the most
``easily'' accessible experimentally are the transverse
spin-correlation ($C_{NN}$) and spin-transfer parameters
($D_{NN}$), which are given by
\begin{eqnarray}
\nonumber
I(pp\!\to\! pK^+\Lambda)\,C_{NN}(pp\!\to\! pK^+\Lambda)
&=&\fourth\mid\! W_{1,s}\!\mid^2\:,\\
\nonumber
I(pn\!\to\! nK^+\Lambda)\,C_{NN}(pn\!\to\! nK^+\Lambda)&=&
\fmn{1}{16}\!\left(\mid\! W_{1,s}\!\mid^2\!-\!\mid\! W_{0,t}\!\mid^2\right)\!,\\
\nonumber
I(pp\!\to\! pK^+\Lambda)\,D_{NN}(pp\!\to\! pK^+\Lambda)
&=&-\half\textrm{Re}(W_{1,s}W_{1,t}^*),\\
\nonumber I(pn\!\to\! nK^+\Lambda)\,D_{NN}(pn\!\to\! nK^+\Lambda)
&= &-\half\textrm{Re}(W_{1,s}W_{1,t}^*). \label{polarise}
\end{eqnarray}
\vspace{-5mm}
\begin{equation}\label{spin_ob}\end{equation}

It follows from these relations that
\begin{eqnarray}
\nonumber%
4I(pn\!\to\! nK^+\Lambda)\left[1+C_{NN}(pn\!\to\!
nK^+\Lambda)\right]\\
=I(pp\!\to\! pK^+\Lambda)\left[1+C_{NN}(pp\!\to\!
pK^+\Lambda)\right]\:,
\end{eqnarray}
so that, in the near-threshold region, the additional measurement
of the spin correlation in $np$ collisions would afford no further
information. Alternatively, measuring just the two spin
correlations would be sufficient to fix \RK{}.
%
%%%%%%%%%%%%%%%%%%%%%%%%%%%%%%%%%%%%%%%%%%%%%%%%%%%%%%%%%%%%%%%%%%%%%%
%
\section{One-boson-exchange models}
\setcounter{equation}{0}

Both strange and non-strange meson exchanges can contribute to
$K^+\Lambda$ production in nucleon-nucleon collisions and the two
sets of diagrams are illustrated on the left and right hand sides
of fig.~1 before the inclusion of effects arising from the
distortion of the initial and final waves.

\begin{figure}[ht]
\begin{center}
\resizebox{0.45\textwidth}{!}{
  \includegraphics{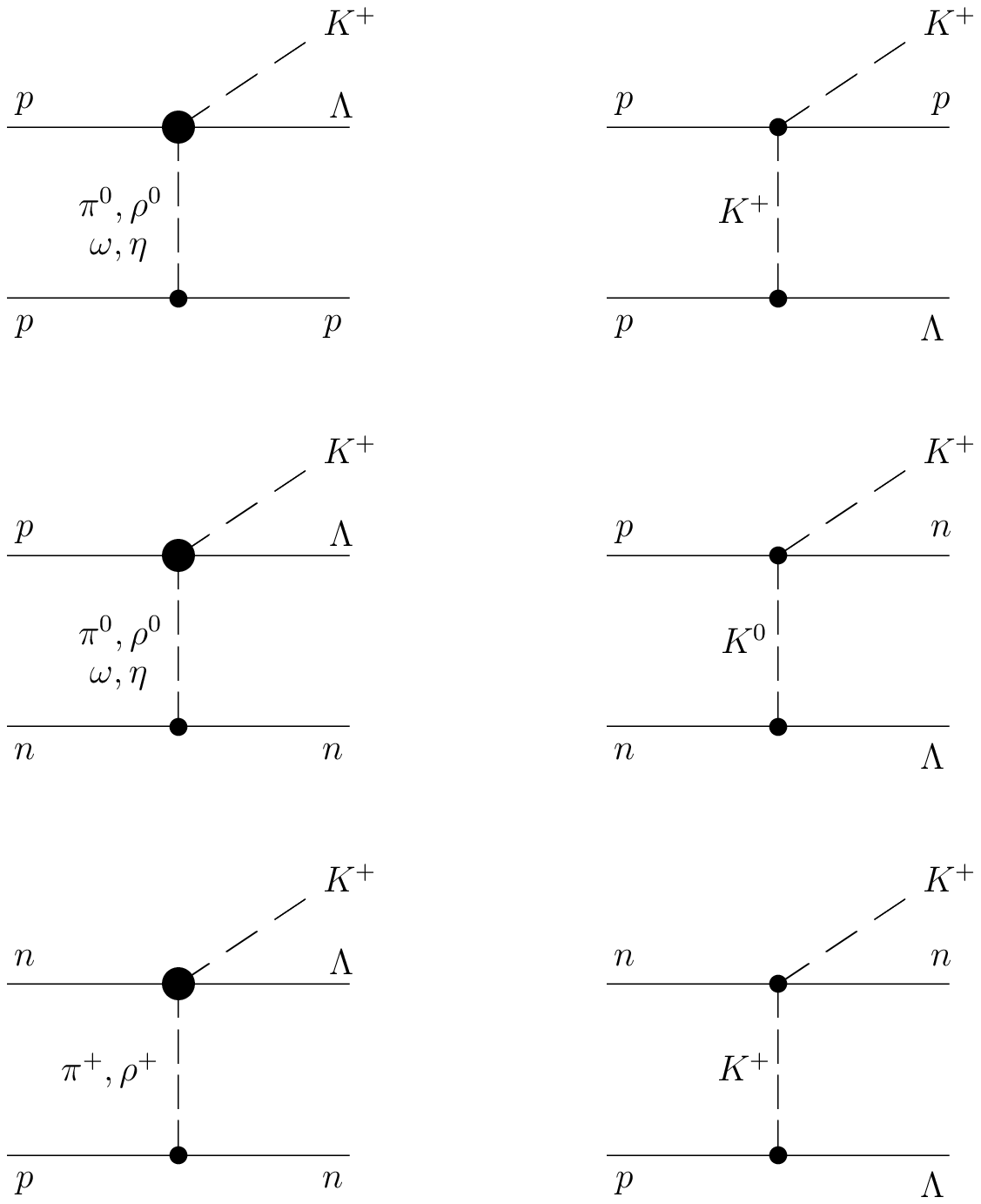}}
 \caption{Bare non-strange and strange one-meson-exchange contributions
 to the $pp\to pK^+\Lambda$ and $pn\to nK^+\Lambda$ amplitudes.}
 \label{fig1}
\end{center}
\end{figure}

Near threshold, the only significant variation is expected to
arise from the spin-singlet and -triplet \fsi\ enhancements. The
relevant propagators, coupling constants, masses \emph{etc}.\ are
evaluated at threshold and so merely contribute to the overall
strength~\cite{FW1}.

Employing the same technique and notation that we used for $\eta$
production, we find that
\begin{eqnarray}
\nonumber W_{1,s}&=& 2\mathcal{B}_{\rho} + 2\mathcal{B}_{\omega}-
\mathcal{D}_{\pi}-\mathcal{D}_{\eta}+\mathcal{D}_K^1\:,\\
\nonumber W_{1,t}&=&
\mathcal{D}_{\pi}+\mathcal{D}_{\eta}+\mathcal{D}_K^1\:,\\
\label{exch_amps}%
W_{0,t}&=& 6 \mathcal{B}_{\rho}-2\mathcal{B}_{\omega}+
3\mathcal{D}_{\pi}-\mathcal{D}_{\eta}+\mathcal{D}_K^0\,,
\end{eqnarray}
where $\mathcal{D}_{\pi,\eta}$ is the amplitude for the exchange
of a pseudoscalar meson and $\mathcal{B}_{\rho,\omega}$ the
dominant vector-exchange term. These amplitudes have the structure
of an $NNx$ coupling constant, the propagator for the meson $x$,
followed by the final $xN\to K^+Y$ transition, which is dominated
by the $S_{11}(1650)$ near threshold. The kaon exchange terms are
similar, except that there is then an $N\Lambda K$ coupling
constant and two isospin possibilities in $K N$ elastic
scattering, leading to the two terms denoted here by
$\mathcal{D}_{K}^{I}$. However, it has been pointed out by
Laget~\cite{Laget} that the isoscalar $K^+N$ scattering is
dominantly $p$-wave and so would contribute relatively little
here.

Though, for completeness, many terms have been included in
eq.~(\ref{exch_amps}), we will concentrate our analysis on just
those for $\pi$ and $\rho$ exchange. The $\eta$ and $K$ terms
might be reduced in importance by the weak coupling constants and,
for $\eta$ production, $\omega$-exchange was reduced rather by the
final transition amplitude.

Using vector dominance to estimate the $\rho N\to \eta N$
amplitudes, we predicted for $\eta$ production that
$\mathcal{D}_{\pi}\approx 0.7\mathcal{B}_{\rho}$~\cite{FW1}, which
led to a reasonable agreement with the large experimental value of
$\mathcal{R}(\eta)$~\cite{Calen}. To estimate the corresponding
value for $K^+$ production, this $\pi/\rho$ factor should be
scaled by the ratio of the amplitudes for the production of $K^+$
with pion and photon beams.
\begin{equation}
\mathcal{D}_{\pi}\approx 0.7 \left(\frac{{\mid\! f(\pi^-p\to
K^0\Lambda)\!\mid^2\: \mid\! f(\gamma p\to \eta p)\!\mid^2}}
{{\mid\! f(\pi^-p\to \eta n)\!\mid^2}\: {\mid\! f(\gamma p\to
K^+\Lambda)\!\mid^2}}\right)^{\!1/2}\, \mathcal{B}_{\rho}\:,
\end{equation}
where we have assumed that the same resonances are responsible for
the production with pions and photons so that, in the absence of
other interactions, the contributions are relatively real.

Taking the experimental data from
refs.~\cite{Binnie,Tiator,Jones,Glander}, we find that
\begin{equation}
\mathcal{D}_{\pi}\approx 0.7\sqrt{\frac{(58\pm 10)(4.6\pm0.2)}
{(810\pm 100)(0.19\pm0.04)}}\,\mathcal{B}_{\rho}=
(0.9\pm0.2)\mathcal{B}_{\rho}\:. \label{ratio2}
\end{equation}
%
%%%%%%%%%%%%%%%%%%%%%%%%%%%%%%%%%%%%%%%%%%%%%%%%%%%%%%%%%%%%%%%%%%%%%%
%
\section{The \bmath{\Lambda N} final-state interaction}
\setcounter{equation}{0}

To determine the overall normalisation of the $pp \to pK^+\Lambda$
cross section, one would need reliable information on the $\Lambda
p$ scattering wave functions, which is still sadly
lacking~\cite{Nijmegen}. However, the shape of the energy
dependence is, to a large extent, fixed by just the $\Lambda p$
scattering lengths and effective ranges in the combination that
gives the positions ($\varepsilon$) of the virtual bound states.
The effect of the \fsi\ on the shape of the cross section can be
included by multiplying the threshold value of $I$ in
eq.~(\ref{sigma}) by the factor~\cite{FW2}
\begin{equation}
\mathcal{Z} = \frac{4}
{\left(1+\sqrt{1+Q/\varepsilon}\right)^2}\:.
\end{equation}
A useful survey of theoretical and experimental information on the
low energy $\Lambda p$ parameters is provided in
ref.~\cite{Frank}. An early experiment~\cite{Alexander} suggested
that the values for the triplet and singlet energies were quite
close ($\varepsilon_t=5.6$~MeV, $\varepsilon_s=5.1$~MeV) and it
has been shown~\cite{Kowina} that the statistical average of these
two ($\varepsilon_t=5.5\pm 0.6$~MeV) gives a good representation
of the $pK^+\Lambda$ total cross section data. Given the current
theoretical uncertainty, for simplicity of presentation, we choose
$\varepsilon_s=\varepsilon_t$.
%
%%%%%%%%%%%%%%%%%%%%%%%%%%%%%%%%%%%%%%%%%%%%%%%%%%%%%%%%%%%%%%%%%%%%%%
%
\section{Results}

\begin{figure}[htb]
\begin{center}
\mbox{\epsfxsize=3.5in \epsfbox{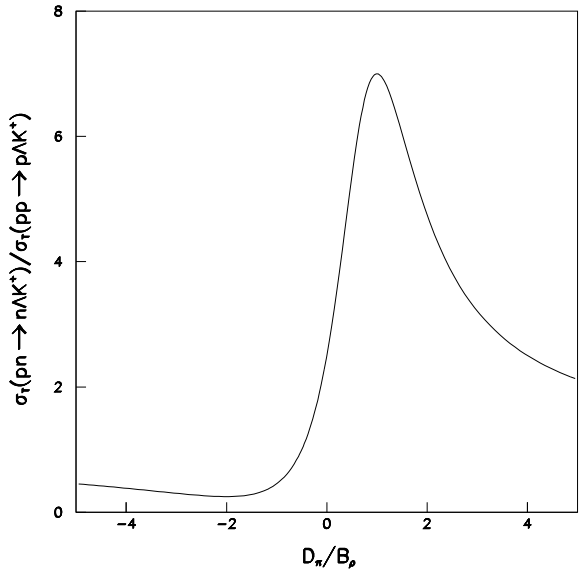}} \caption{Prediction
for \RK\ as a function of $\mathcal{D}_{\pi}/\mathcal{B}_{\rho}$,
assumed to be real. Taking this value from Eq.~(\ref{ratio2})
leads to a ratio close to the maximum of 7. \label{fig_2}}
\end{center}
\end{figure}

By taking $\varepsilon_s=\varepsilon_t$, it follows that \RK\
should not depend upon the excitation energy $Q$ and in
Fig.~\ref{fig_2} the prediction for this has been drawn as a
function of $x=\mathcal{D}_{\pi}/\mathcal{B}_{\rho}$, where it has
been assumed that the $\pi$- and $\rho$-exchange amplitudes have
the same phase.

\input epsf

As $x\to\infty$, $\RK\to 1$~\cite{AG}, which is very different to
the factor of five expected for $\eta$ production under similar
assumptions. This difference arises, in part, because we do not
include the contribution of the $K^0$ in the definition of
$\mathcal{R}$, but also because the $W_{1,t}$ term is forbidden by
the Pauli principle for the $pp\to pp\eta$ reaction. On the other
hand, pure $\rho$ exchange leads to $\RK=2.5$ and for a wide range
of $x$ the ratio is well over unity. It is interesting to note
that the figure of $0.9$, resulting from the crude scaling model
of eq.~(\ref{ratio2}), corresponds almost to the peak value of 7
shown in Fig.~\ref{fig_2}.

\begin{figure}[htb]
\begin{center}
\mbox{\epsfxsize=3.5in \epsfbox{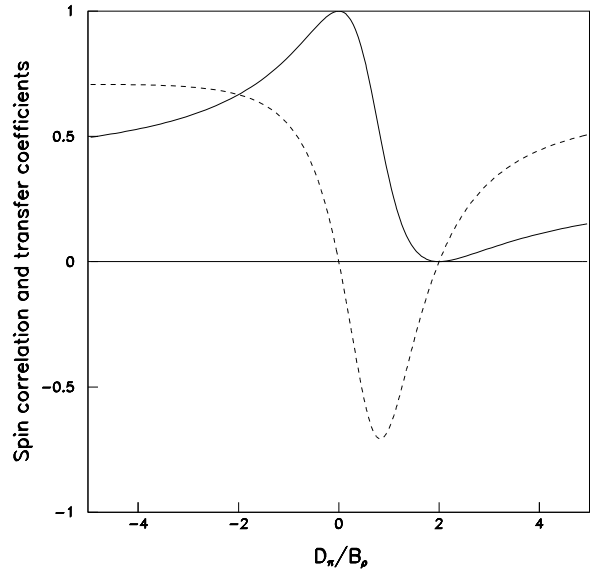}} \caption{Transverse
spin-correlation $C_{NN}$ (solid curve) and spin-transfer $D_{NN}$
(broken curve) parameters  for the $pp\to pK^+\Lambda$ reaction as
functions of $\mathcal{D}_{\pi}/\mathcal{B}_{\rho}$, assumed to be
real. Taking the value of this from Eq.~(\ref{ratio2}) leads to a
large negative $D_{NN}$, without invoking kaon
exchange~\cite{Laget}. \label{fig_3}}
\end{center}
\end{figure}

From eqs.~(\ref{polarise}) and (\ref{exch_amps}), it is seen that,
with this value of $x$, one expects $C_{NN}(pp\!\to\!pK^+\Lambda)
= 0.43$, to be contrasted with the $0.5$ and $1.0$ expected for
pure $\pi$ and $\rho$ exchange respectively. The variation of both
$C_{NN}$ and $D_{NN}$ with $x$ is shown in fig.~\ref{fig_3}, where
it is seen that $D_{NN}$ has an even more interesting behaviour,
with a minimum of about $-0.7$ for $x=0.9$. This is to be
contrasted with the $+2/3$ and $0$ expected for pure $\pi$ and
$\rho$ exchange respectively.
\input epsf
%
%%%%%%%%%%%%%%%%%%%%%%%%%%%%%%%%%%%%%%%%%%%%%%%%%%%%%%%%%%%%%%%%%%%%%%
%
\section{Conclusions}

We have studied the charge and spin dependence of the $pp\to
pK^+\Lambda$ total cross section near threshold in the region
where the final particles are in relative $S$-states. The overall
cross section strength is hard to estimate with any confidence,
due principally to the poor knowledge of the $\Lambda N$
potential. Scaling the cross section from that for $\eta$
production using the Bargmann potential, as in ref.~\cite{FW2},
avoids some of the problems associated with initial state
interactions~\cite{Speth2} and gives a good energy dependence.
However, it leads to an estimate that is too low by a factor of up
to five. This probably indicates that the short-range part of the
$\Lambda p$ interaction is less repulsive than that for
$pp$~\cite{Nijmegen}.

Since there are only three amplitudes describing $K^+\Lambda$
production on the proton and neutron near threshold, it is clear
that there should be some model-independent relations between the
charge and spin dependence of the observables. To go further than
this, we have worked in a simplified meson-exchange model, where
the amplitudes have been assumed to be in phase. Keeping only the
$\pi$ and $\rho$ terms, and scaling their relative strength from
that found for $\eta$ production, we find that production of
$K^+\Lambda$ on the neutron could indeed be much stronger than on
the proton. However, the prediction does depend upon cancellations
and is far less robust than that in the $\eta$ case.

The spin-correlation and transfer parameters are also expected to
depend sensitively upon $x$, the relative $\pi/\rho$ strength. Of
especial interest is $D_{NN}$ which, though +2/3 for $\pi$
exchange and $0$ for $\rho$ exchange, is predicted to be strongly
negative for our preferred value of $x$, though this does depend
upon our phase \emph{ansatz}. The negative value found for
$D_{NN}$ in $pp\to pK^+\Lambda$ in different kinematic conditions
away from threshold by the DISTO group~\cite{DISTO} was taken as
evidence for the dominance of kaon over pion
exchange~\cite{Laget}, but it is important to stress that the
possibility of $\rho$ exchange was not considered by these
authors.

We have only looked in detail at the consequences of $\pi$ and
$\rho$ exchanges. Whether other terms are significant or not might
be determined from the spin observables of eq.~(\ref{spin_ob}). In
particular, the measurement of $C_{NN}$ in $pp$ collisions as well
as of the $pn$ and $pp$ cross section would allow one to separate
the magnitudes of the $W$ amplitudes. Since this would fix both
the $\pi$ and $\rho$ amplitudes and their interference, it would
then predict unambiguously the spin-transfer coefficient $D_{NN}$
and this would allow a test of the $\pi/\rho$ model.

\begin{acknowledgement}
This work was initiated through discussions with V.~Koptev and
Y.~Valdau during a Royal Society sponsored visit by one of the
authors (CW) to the Petersburg Nuclear Physics Institute. The
authors are very grateful for constructive comments from
C.~Hanhart and A.~Gasparyan.
\end{acknowledgement}

%
%%%%%%%%%%%%%%%%%%%%%%%%%%%%%%%%%%%%%%%%%%%%%%%%%%%%%%%%%%%%%%%%%%%%%%%%%%%
%

\end{document}